\documentclass[iop]{emulateapj} 

\usepackage{amssymb,amsmath}
\usepackage{graphicx}
\usepackage{multirow}
\usepackage{natbib}

\usepackage{url}
\usepackage{float}
\usepackage{bm}
\footnotesize
\newdimen\minuswidth    
\setbox0=\hbox{$-$}
\minuswidth=\wd0
\catcode`@=\active
\def@{\kern\minuswidth}
\newdimen\digitwidth    
\setbox0=\hbox{\rm0}
\digitwidth=\wd0
\catcode`!=\active
\def!{\kern\digitwidth}
\normalsize

\usepackage[usenames,dvipsnames]{color}

\footnotesize
\newdimen\digitwidth    
\setbox0=\hbox{\rm0}
\digitwidth=\wd0
\catcode`!=\active
\def!{\kern\digitwidth}
\normalsize

\begin{document}

%

\title{Pulsar Observations of Extreme Scattering Events }
\author
{W.~A.~Coles\altaffilmark{1}, 
M.~Kerr\altaffilmark{2}, 
R.~M.~Shannon\altaffilmark{2}, 
G.~B.~Hobbs\altaffilmark{2}, 
R.~N.~Manchester\altaffilmark{2},
X.-P.~You\altaffilmark{3},
M. Bailes\altaffilmark{4},
N. D. R. Bhat\altaffilmark{5},
S. Burke-Spolaor\altaffilmark{6},
S. Dai\altaffilmark{7,2},
M. J. Keith\altaffilmark{8},
Y. Levin\altaffilmark{9},
S. Os{\l}owski\altaffilmark{11,10},
V. Ravi\altaffilmark{12,2},
D. Reardon\altaffilmark{9,2},
L. Toomey\altaffilmark{2},
W. van Straten\altaffilmark{4},
J. B. Wang\altaffilmark{13,14},
L. Wen\altaffilmark{15},
X. J. Zhu\altaffilmark{15,2}
}
\altaffiltext{1}{ECE Dept., University of California at San Diego, La Jolla, CA, 92093-0407, U.S.A; bcoles@ucsd.edu} 
\altaffiltext{2}{ATNF, CSIRO Astronomy \& Space Science, P.O. Box 76, Epping, NSW 1710, Australia}
\altaffiltext{3}{Southwest University, Chongqing, China}
\altaffiltext{4}{Centre for Astrophysics and Supercomputing, Swinburne University of Technology, P.O. Box 218, Hawthorn, VIC 3122}
\altaffiltext{5}{International Centre for Radio Astronomy Research, Curtin University, Bentley, WA 6102, Australia}
\altaffiltext{6}{California Institute of Technology, Pasadena, 1200 E California Blvd, CA 91125, U.S.A.}
\altaffiltext{7}{Department of Astronomy, School of Physics, Peking University, Beijing 100871, China}
\altaffiltext{8}{Jodrell Bank Centre for Astrophysics, School of Physics and Astronomy, The University of Manchester, Manchester M13 9PL, UK}
\altaffiltext{9}{Monash Center for Astrophysics, School of Physics and Astronomy, Monash University, Vic 3800, Australia }
\altaffiltext{10}{Max-Planck-Institut f{\"u}r Radioastronomie, Auf dem H{\"u}gel 69, D-53121 Bonn, Germany}
\altaffiltext{11}{Department of Physics, Universit{\"a}t Bielefeld Universit{\"a}tsstr. 25 D-33615 Bielefeld, Germany}
\altaffiltext{12}{School of Physics, University of Melbourne, Vic 3010, Australia }
\altaffiltext{13}{Xinjiang Astronomical Observatory, Chinese Academy of Science, 150 Science 1-Street, Urumqi, Xinjiang, China, 830011}
\altaffiltext{14}{University of Chinese Academy of Sciences, Beijing, China, 100049 }
\altaffiltext{15}{University of Western Australia, 35 Stirling Hwy, Crawley, WA 6009, Australia}

\begin{abstract}
Extreme scattering events (ESEs) in the interstellar medium (ISM) were first observed in regular flux measurements of compact extragalactic sources. They are characterized by a flux variation over a period of weeks, suggesting the passage of a ``diverging plasma lens'' across the line of sight. Modeling the refraction of such a lens indicates that the structure size must be of order AU and the electron density of order 10s of cm$^{-3}$. Similar structures have been observed in measurements of pulsar intensity scintillation and group delay. Here we report observations of two ESEs showing increases in both intensity scintillation and dispersion made with the Parkes Pulsar Timing Array (PPTA). These allow us to make more complete models of the ESE, including an estimate of the ``outer-scale'' of the turbulence in the plasma lens. 
These observations show clearly that the ESE structure is fully turbulent on an AU scale. They provide some support for the idea that the structures are extended along the line of sight, such as would be the case for a scattering shell.
The dispersion measurements also show a variety of AU scale structures which would not be called ESEs, yet involve electron density variations typical of ESEs and likely have the same origin.
\end{abstract}

\section{Introduction}
The original observation of ESEs (Fiedler et al, 1987), shown in Figure 1, suggests the passage of a diverging lens across the line of sight, as the flux is weaker in the middle of the event and piles up towards the edge. A convex ``blob'' of high density plasma would provide such a lens. The refractive index variations would be smaller at a higher observing frequency and the observations are consistent with this behavior, so it has been assumed that this is the basic mechanism causing the ESE event. 

The observations attracted wide attention and many reports of similar structures in the ISM have been published. These reports include pulsar observations of correlated fluctuations in group delay and flux (Cognard et al, 1993, Lestrade et al, 1998), persistent phase gradients (Gupta et al, 1994), enhanced diffractive scintillation (Stinebring et al, 2001; Hill et al, 2005; Brisken et al., 2010); enhanced angular broadening (Lazio et al, 2000) and increased dispersion (Keith et al, 2013). However observations of both enhanced diffractive scattering and dispersion have not previously been reported for the same event. The combination allows one to make a more thorough analysis of the ESE including an estimate of the ``outer scale'' of the turbulence in the plasma. It has become common to use the term ESE for both the event and the plasma blob responsible for it. We will follow this convention.

Pulsar timing arrays are designed to make precise measurements of the group delay of an array of pulsars every few weeks (see, for instance, Manchester et al. 2013 and references therein) for decades. The primary objective of PTAs is the direct detection of gravitational waves with periods of the order of a decade and for many pulsars the primary noise source is fluctuations in the electron density of the ISM. To correct for the ISM noise, PTAs also make precise measurements of the dispersion in the ISM every few weeks. These observations provide an excellent window of opportunity to view fluctuations of the ISM on a scale of AU in general, and to study ESEs in particular.

\begin{figure}
\centerline{
\includegraphics[width=8 cm]{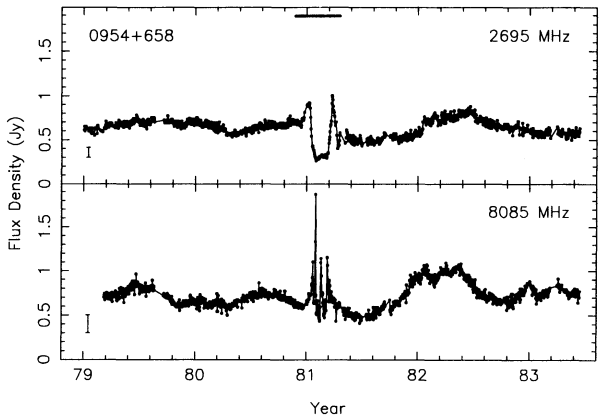}}
\caption{
\label{fig1}
Original ESE observations of Fiedler et al (1987).}
\end{figure}

Pulsar observations are made with filter-banks because the change in dispersive delay over a typical observing band is much greater than the pulse width or even the pulse period, so the different frequency channels must be aligned before they can be averaged over the observing band. This provides a good measurement of the total electron column density, but for PTA purposes it must be supplemented by observations in two different bands separated more widely than any existing receiver bandwidth. Inter-band measurements of millisecond pulsars can achieve accuracies of 1:10$^5$ in dispersion. 

Filter-bank observations are also ideal for measuring the diffractive scintillation because the ``dynamic spectrum'' of intensity scintillations is a two-dimensional random process with characteristic ``scales'' in both frequency, $\nu_0$, and time, $\tau_0$.
These scales depend on the strength of scattering, the distance to the pulsar, the velocity of the pulsar and the location of the scattering region. The distance and proper motion of the pulsar can often be obtained from the timing model. This is an astrometric model which is fitted to the time of arrival (ToA) of the pulses. For nearby pulsars the distance can be determined directly from the parallax, which causes a biannual sine wave in the ToAs. Otherwise the distance can be inferred from the measured dispersion using a model of the Galactic electron density (Taylor and Cordes, 1993, Cordes and Lazio, 2002). The location of the screen and the strength of scattering can then be estimated from the measured $\nu_0$ and $\tau_0$.

With measurements of the electron column density (DM) from the dispersion, and the level of scattering from $\nu_0$, and $\tau_0$, we can model the turbulence in the observed ESEs and compare this model with the ``average'' turbulence in the entire line of sight. In a general sense, it is surprising that one can detect structures in column  density or scattering which are as small as an AU in a line of sight that is 100s of pc long. The exceptional accuracy of DM measurement provided by inter-band measurements make this possible. Scattering measurements cannot be made with accuracy greater than $\sim$\,20\%, but we will show that AU structures can cause exceptionally strong scattering if they are fully turbulent, i.e. if the fluctuations in electron density $n_e$ are comparable with the mean $n_e$ on an AU spatial scale. 

\section{Scattering in the ISM}
A radio wave propagating through an ESE in the ISM will be scattered into an angular spectrum of plane waves $B(\vec{\theta})$ by fluctuations in $n_e$ transverse to the line of sight. The $n_e$ fluctuations cause only phase changes in the radio wave, but as the scattered waves propagate to the Earth they interfere, causing intensity scintillations to build up with distance from the ESE. When the scintillations are weak, i.e. the intensity fluctuations are small and the bandwidth is broad, the intensity fluctuations will have a single characteristic scale which is of order of the Fresnel scale. As the scintillation gets stronger the characteristic scale bifurcates into a diffractive scale, which decreases and a refractive scale which increases. The geometric mean of the two scales remains the same as the weak scattering scale. The diffractive scintillation causes a pulse delay $\tau_{del}$ which depends on $\theta$. This $\tau_{del}$ will de-correlate the intensity fluctuations at different wavelengths, giving a coherence bandwidth $\nu_0 = 1/2 \pi \langle \tau_{del}\rangle$. The refractive scintillations remain relatively broadly correlated over wavelength.
Scintillations are normally observed as a time series, as the ISM drifts across the line of sight with a transverse velocity $\vec{V}_{\rm{eff}}$. 
For the observations discussed here, and for typical pulsars at frequencies around 1 GHz,
the diffractive scale is of order 10 min and the refractive scale is of order 1 day. In our observations, of the order of an hour, one sees diffractive scintillations only, but the mean flux from one day to the next will vary due to the refractive scintillations. For a discussion of scattering see Coles et al. (2010) and for application to the ISM see Rickett (1990).

\subsection{Diffractive Intensity Scintillation}

Here we will refer to the scattering plasma as a generalized ``screen'', which may be an ESE or simply a slab of ISM. When the pulsar is at a distance L and the screen is at distance $\zeta L$ from the pulsar we can write $V_{\rm{eff}}$ as
\begin{equation}
\vec{V}_{\rm{eff}} = \vec{V}_E \zeta + \vec{V}_P (1-\zeta) - \vec{V}_{ISM}.
\end{equation}
Here all velocities are projected onto the celestial sphere.
$\tau_{del}$ can be written as a function of $\vec{\theta}$ given the screen location. A wave scattered at angle $\theta$ will leave the pulsar at angle $\theta_P = (1 - \zeta) \theta$ and will arrive at the Earth at angle $\theta_E = \zeta \theta$.
\begin{equation}
\tau_{del} = \zeta ( 1 - \zeta) \theta^2 L / 2c
\end{equation}

If the phase fluctuations in the screen have stationary gaussian differences the autocovariance of the field at the screen 
$\rho_e (\vec{r}) = \langle e( \vec{r}') e^*( \vec{r}' + \vec{r}) \rangle$ is related to the structure function of the screen phase $D_{\phi} (r) = \langle (\phi (r') - \phi(r+r' ) )^2\rangle$ by
$\rho_e (\vec{r}) = \exp(-0.5 D_{\phi}(\vec{r}))$. If the turbulence is Kolmogorov
\begin{align}
D_{\phi} (r) &= (r / s_0 )^{5/3} \mbox{\ \ \ \, for\ } r < s_{out} \mbox{\ \ and} \\ 
	&= (s_{out} / s_0 )^{5/3} \mbox{\  for\ } r > s_{out}. \nonumber
\end{align}

The brightness distribution $B(\vec{\theta})$ is the Fourier transform of $\rho_e (\vec{r})$,
\begin{equation}
B(\vec{\theta}) = \iint \rho_e (\vec{r}) \ exp(-j 2 \pi \vec{r} \cdot \vec{\theta} / \lambda )\  d^2 r
\end{equation}
Both $\rho_e (\vec{r})$ and $B(\vec{\theta})$ are almost gaussian,
so if we define
the $1/\sqrt{e}$ width of $\rho_e$ as $s_0$ then the $1/\sqrt{e}$ width of $B(\theta)$ is $\theta_0 = \lambda / 2 \pi s_0$, and
$D_{\phi} (s_0 ) = 1$. 

The bandwidth of the intensity fluctuations is then $\nu_0 = 1 / 2 \pi \tau_{del} (\theta_0)$. All the scintillation parameters, including $\nu_0$ and $\tau_{del}$,
are slowly varying functions of frequency. We can define them as approximately constant over a narrow band centered on $\nu_M$.

If $\nu_0 < \nu_M$ the scattering is strong and
\begin{equation}
\frac{\nu_0}{\nu_M} = \frac{2}{\zeta (1-\zeta)} \left(\frac{s_0}{r_f} \right)^2,
\end{equation}
where $r_f = \sqrt {\lambda L / 2 \pi}$ is the Fresnel scale at $\nu_M$. This is the case for all the observations in the PPTA except for those of PSR J0437-4715. 

The intensity time scale $\tau_0$ (at 1/e) is defined by
$s_0 = V_{\rm{eff}}\, \tau_0$, so with measurements of $\nu_0$ and $\tau_0$ and knowledge of L, 
one can find the location of the screen $\zeta$ and the spatial scale $s_0$. 
If we neglect $V_E$ and $V_{ISM}$, which is often reasonable, we obtain 
\begin{align}
\zeta/(1-\zeta) &= (V_P \tau_0 / r_f )^2 (2 \nu_M / \nu_0 ) \mbox{\ \ \  and} \\
s_0 &= V_P (1-\zeta) \tau_0. 
\end{align}
If $V_E$ is important one must add $\vec{V}_E \zeta / (1-\zeta)$ to $\vec{V}_P$ and the solution can be obtained iteratively.

\subsection{Outer Scale Model}

The outer scale of any turbulent system is difficult to measure but observations suggest that in most cases the outer scale is comparable with the dimensions of the system.
Thus the outer scale $s_{out}$ of an ESE should be comparable with the smallest dimension of the ESE.
The phase structure function limit for $r \ge s_{out}$ is equal to twice the phase variance, $D_{\phi} (s_{out}) \approx 2 \langle \phi ^2 \rangle$ by definition. If the path length $W_z$ in the scattering structure equals $s_{out}$ then $\langle \phi ^2 \rangle = \langle \phi \rangle^2$ because the rms density equals the mean density. This can be used to relate $\langle \phi ^2 \rangle$ to $\langle \phi  \rangle$ for thicker screens in which $W_z = N s_{out}$. In this case $\langle \phi ^2 \rangle = \langle \phi \rangle^2 / N$.
A spherical ESE would limit at $D_{\phi} (W_z) = 2 \langle \phi \rangle^2$. An ESE extended along the line of sight, such as a shell model, would have
$s_{out} \approx W_t$ the transverse width. The structure function would saturate at
$D_{\phi} (W_t) = 2 \langle \phi \rangle^2 (W_t / W_z)$.

Observationally we can measure $\langle \phi \rangle$ for the ESE through the change in group delay, i.e. $\langle \phi \rangle = 2 \pi \nu_M \delta t_{grp}$ and  $s_0$ from the intensity scintillation. We can also measure $W_t$ from the time it takes to cross the line of sight, but we don't know $W_z$.
If $W_t \le W_z$ then $s_{out} = W_t$ and
\begin{equation}
(W_t / s_0 )^{5/3} = 2 (2 \pi \nu_M \delta t_{grp} )^2 (W_t / W_z).
\end{equation}
We can solve this expression for $W_z$, which will be valid only if $W_z \ge W_t$. Otherwise one would have to set $s_{out} = W_z$ and solve
\begin{equation}
(W_z / s_0 )^{5/3} = 2 (2 \pi \nu_M \delta t_{grp} )^2 
\end{equation}
for $W_z$. The latter case did not occur in our observations. 

This analysis can also be applied to the entire line of sight, excluding the ESE. The scattering would be modeled as the superposition of N independent slabs of depth $s_{out}$ distributed randomly over the line of sight. The $\zeta$ estimated would be a weighted average of the slab locations.
In this case we derive $t_{grp}$ from the mean DM and solve 
\begin{equation}
(s_{out}/ s_0 )^{5/3} = 2 (2 \pi \nu_M t_{grp} )^2 (s_{out} / L).
\end{equation}
for an average $s_{out}$. Such an analysis neglects a number of biassing effects and could be expected only to provide the correct order of magnitude for $s_{out}$.

\section{Data Analysis}

The dynamic spectra were measured with standard procedures in \textsc{psrchive} (Hotan, van Straten \& Manchester 2004) which provide a flux estimate by fitting a template to the measured pulse profile. This is done for each frequency channel, m = 1 to M and each sub-integration n = 1 to N. The sub-integration time was 1 min and the observing time was usually 64 min. The auto-covariance $C(\tau',\nu')$ of the dynamic spectrum $S(\tau, \nu)$ is estimated using
\begin{equation}
C(n',m') = \frac{1}{NM} \sum \sum S(n,m) S(n+n', m+m').
\end{equation}
This estimator multiplies the covariance by a triangle function but keeps the errors roughly independent of lag. We eliminate the bias caused by this  triangle by including the triangle in the model fit as shown below.
This temporal model is well supported theoretically, but the frequency model is only a rough approximation. A gaussian has been used but it is not a good approximation. A Lorentzian has also been used but it does not approximate the actual curve as well as an exponential near the origin. Both the exponential and the Lorentzian do well at larger lags. We did not use a more complex model because the errors caused by weakness in the model are dwarfed by the actual variations.

We fit the auto-covariance to the theoretical model, parameterized by the time and frequency scales, $\tau_0$ and $\nu_0$ and the variances A (white noise) and B (scintillation).
\begin{align}
C(0,0) &= A + B  \\
C( n' ,0) &= B \exp (-| \tau(n') / \tau_0 |^{5/3}) (N-n')/N\\
C(0, m' ) &=  B \exp (-|\nu(m') / \nu_0 |) (M-m')/M
\end{align}
For all the PPTA observations the time scale $\tau_0 \gg 1$ min, so the white noise delta function (A) is clearly separated from the temporal scintillation. However the bandwidth $\nu_o$ is often near the frequency resolution (channel width) of the receiver and the white noise delta function is not always clearly separated from the scintillation. So we fit for A, B and $\tau_0$ in the temporal covariance, equation (12), first, then we fit for  $\nu_0$ in the frequency covariance, equation (13), holding A and B fixed. 

\section{Observations}

The fluctuations in DM, $\delta DM(t)$, were determined with the technique discussed in Keith et al (2013). The fluctuations can be measured with a precision of the order of 1:10$^5$, which is much higher than the absolute accuracy with which DM can be measured. The problem with absolute DM is that the pulse shape changes with frequency and these changes make it impossible to compare the group delay at different frequencies precisely. However the pulse shape is remarkably stable with time. Thus we can measure very small time variations even though we do not know the absolute DM with comparable accuracy. Here we show $\delta DM(t)$ with respect to zero and provide the mean DM separately.

The distance of the pulsar is not known precisely in either of the two ESE observations. This uncertainty propagates into the velocity because, although the angular proper motion is well determined from the timing model, it must be multiplied by the distance to obtain $V_P$. In both cases the distance computed from the Taylor and Cordes (1993) model differs significantly from that computed using the newer Cordes and Lazio (2001) model. Accordingly we show the scattering models derived from both distance estimates to illustrate the sensitivity of the analysis to distance.

\subsection{ESEs}

One of the ESEs, in observations of J1603$-$7202, is in the PPTA data release 1 (Manchester et al. 2013) and the dispersion variations were discussed by Keith et al. (2013).  The other ESE, in observations of J1017$-$7156, was found in the HTRU survey and the dispersion variations were noted by Ng et al, (2014). It was included
in continuing PPTA observations, as yet unpublished, but taken with the same instruments and the same primary analysis procedures as discussed in these earlier papers. 

The observations for J1603-7202 are shown in Figure 2.
\begin{figure}
\includegraphics[width=9.2cm]{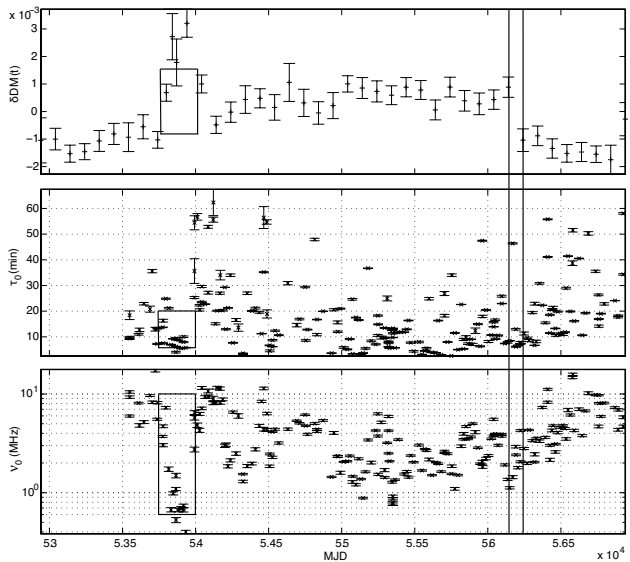}
\caption{
\label{fig2}
Observations of J1603$-$7202. The ESE between 53740 and 54000 is marked by boxes indicating the parameters used in the analysis. Two vertical lines align the scattering observations with an obvious step in the DM(t).}
\end{figure}
The ESE between MJDs  53740 and 54000 is clear in all three parameters, $\delta DM$, $\tau_0$, and $\nu_0$, but best defined in $\nu_0$. The ESE parameters, shown as a solid box in the graph, were fit by eye because neither the observations, nor the analysis are precise enough to justify a model fit. The pulsar parameters, the ESE measurements, and the derived parameters are given in Table 1 labeled (ese).
Ignoring $V_E$ we find $0.39 < \zeta < 0.46$, i.e. the ESE is roughly midway between the pulsar and the Earth. 
If $s_{out} = W_t$, we find $W_z = 29 W_t$, and $3.4 < \langle n_e\rangle < 4.1$ cm$^{-3}$.
This lends support to shell type models including the corrugated reconnection sheet of Pen and Levin (2014). 
It is also possible that the ESE is spherical ($W_t = W_z = s_{out}$) but not fully turbulent with rms electron density $\approx$ 20\% of the mean density. In this case $\langle n_e\rangle \approx 99$ cm$^{-3}$. This $n_e$ is so high that we searched for other observations, such as H$\alpha$, that might show it. We did not find anything interesting, but the ESE is so small that it might require a targeted observation. At this time we regard the spherical ESE model as less likely than the extended model.
\begin{table*}
\caption{\label{rmss}Model parameters for ESEs.}
\begin{tabular}{lccccccccrccrcc}
\hline
PSR & $L$  &   $V_P$ &  DM*                & $\delta_{DM}$*  &$\tau_{DM}$ & $\nu_0$ & $\tau_0$ & $\zeta$ & $s_0$ & $W_t$ & $W_z$ & $t_{grp}$ & $s_{out}$ & $\langle n_e\rangle$\\
         & pc    &    km/s   &   &     & days             & MHz &       min         &              &    km   &   AU   &  AU  &    &     &   cm$^{-3}$ \\
\hline
\hline
J1603$-$7202(ese) & 1640$^1$ & 61 & 38  &  0.0023 & 260 & 0.6 & 5  & 0.46 & 9805 &  4.9 & 142 & 5.1$\mu$s & 4.9AU & 3.4 \\
			    & 1170$^2$ &  44 &       &             &         &      &       & 0.39 & 8089& 4.0 & 117 & & 4.0AU & 4.1 \\
J1603$-$7202(los) & 1640 & 61  & 38  &  -           & -      & 10 & 20  & 0.45 & 39960 &  -    & -        & 84ms & 1.0pc & - \\
			    & 1170 &  44 &       &             &         &      &       & 0.38 & 32684& -     & -         &          & 1.0pc & - \\
J1017$-$7156(ese) & 3000$^2$ & 144& 94 & 0.0015 & 200 & 2.0 & 2.5 & 0.17 & 18030 & 13.9 & 83 & 3.3$\mu$s & 13.9AU & 3.7 \\
			  & 8000$^1$ & 384&    &       &     &     &     & 0.35 & 37698 & 29 & 174 &          & 29AU & 1.8 \\
J1017$-$7156(los) & 3000 & 144& 94 &      -       &  -     & 10 & 10 & 0.39 & 52892 &   -      &  -      & 208ms & 12pc & - \\
     & 8000 & 384&    &      -       &  -     &    &   & 0.63 & 85671 &   -      &  -      &     & 9.1pc & - \\
\hline
\end{tabular}

DM$^*$ in pc cm$^{-3}$;  L$^1$  by Taylor and Cordes (1993) DM model; L$^2$ by Cordes and Lazio (2002) DM model.
\end{table*}

We have applied equation 9 to the full line of sight, exclusive of the ESE. The results are in Table 1 labeled (los).
Ignoring $V_E$ we find $0.38 < \zeta < 0.45$ and
the resulting average outer scale for the entire line of sight is $s_{out} = 1.0$ pc. This is consistent with the work of Haverkorn et al, (2004, 2008).

The variation in $\tau_0$ is partially due to significant changes in $V_{\rm eff}$ caused by a combination of $V_E$ and the orbital velocity of the binary system. The variation in $\nu_0$ is not affected by the velocity and must be due to real variations in the turbulence level of the ISM. These are very substantial variations and add to the accumulating evidence that the ISM is far from homogeneous on AU scales.

The observations for J1017$-$7156 are shown in Figure 3. Here the ESE between MJDs 55650 and 55800 is less well-defined than the one in J1603$-$7202, but the recovery step after the ESE is very abrupt in $\tau_0$ and $\nu_0$. The parameters of the analysis are given in Table 1. In this case we are inclined to put more weight on the Cordes and Lazio (2002) Galactic model because it is newer and the $V_P$ calculated with the Taylor and Cordes (1993) model is exceptionally high. 
Ignoring $V_E$ we find $\zeta < 0.17$, i.e. the ESE is closer to the pulsar than the Earth. 
If $s_{out} = W_t$, we find $W_z = 6 W_t$, so $\langle n_e\rangle = 3.7$ cm$^{-3}$. Again a shell or corrugated reconnection sheet would be favored, but a spherical model is tenable if the rms electron density is less than the mean by a factor $\approx \sqrt{5}$. In this case $\langle n_e\rangle = 22$ cm$^{-3}$.

We have applied equation 9 to the full line of sight, exclusive of the ESE. The results are shown in Table 1.
Ignoring $V_E$ we find $s = 0.39$ and 
the resulting ``effective'' outer scale for the entire line of sight is $s_{out} = 12$ pc. This is somewhat larger than expected but not inconsistent with the work of Haverkorn et al, (2004, 2008).

One can see that $\nu_0$ is less variable in this pulsar. The bandwidth is slightly smaller which provides more degrees of freedom in the covariance estimate, but not enough to explain the significant reduction in variability. This must be a difference in the ISM itself. 
There is also an indication of a very short ESE near MJD 56660. Since this possible event is defined only by a couple of samples we will not analyze it further.

These calculations are only accurate to first order. In particular the parameters of the ESE are approximate and the DM distance is quite uncertain because of uncertainties in the Galactic $n_e$ models. The calculations could be improved by including $V_E$, the binary velocity of the pulsar, the local velocity of the ISM, and the anisotropy of the turbulence as was done for the double pulsar J0737$-$3039A/B (Rickett et al, 2013). We hope to do such an analysis for several PPTA pulsars, but it is not necessary to obtain a first order model of the ESEs reported here. Our goal is to establish that the DM and diffractive scattering observations are consistent with a simple model of the ESE and provide a good estimate of the outer scale of the turbulence in the ESE.

\begin{figure}
\includegraphics[width=8.2cm]{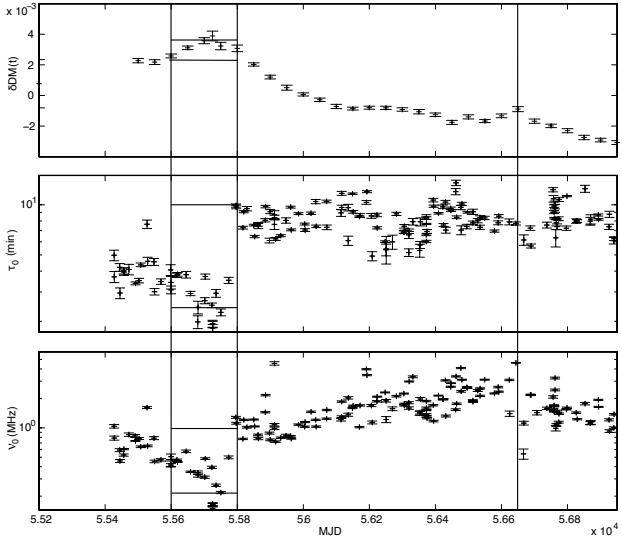}
\caption{
\label{fig3}
Observations of J1017$-$7156. The ESE between 55600 and 55800 is analyzed. A smaller event at 56640 appears to be real but it not analyzed in detail.}
\end{figure}

\subsection{Other Events}
One can see an abrupt step in DM(t) for J1603$-$7202 in Figure 2 at MJD = 56200, which not clearly present in $\tau_0$ or in $\nu_0$.  We have double-checked the observations and see no reason to believe that this step is spurious. Both $\tau_0$ and $\nu_0$ are extremely variable so perhaps a correlation is simply buried in the natural variation. Nevertheless the integrated electron density changes by -0.002 pc cm$^{-3}$ in less than 75 days or 2.6 AU. Clearly the local density change must be of the same order as that of the earlier ESE. Since the density decreases the earlier identified ESE must be part of a more extended high density ``local cloud''.
Although there are no other statistically significant changes in DM(t) there are significant changes in the bandwidth, which can only be due to changes in the turbulence level.

A ``pinhole'' in the ISM can be seen in observations of J1713+0747 shown in Figure 4. This pinhole was evident in the observations of Keith et al. (2013) but was not discussed because it consists of a single DM(t) estimate. However it must be real because it is also seen in observations from Arecibo, Greenbank and Nancay which will be discussed in a paper in preparation (Lentati, private communication).
An under-dense region corresponds to a converging ``lens'', so one might expect some focussing. Accordingly we have also displayed the flux density on Figure 4. The problem of seeing small flux changes in the presence of strong diffractive scintillation is particularly clear here. However there is weak evidence for a very short jump in $\nu_0$ and $\tau_0$ at the DM(t) drop of 0.01 pc cm$^{-3}$, which would indicate that lower scattering accompanied the lower density. Of course density cannot go negative, so such a small pinhole must be an ESE with a hole in it. The observations in fact show a marginally significant (2$\sigma$) jump, followed by the more significant drop and recovery to the pre-event level. A shell expanding across the line of sight faster than the Alfv\'{e}n speed would have such an effect. Such events are very common in the solar wind, they push up a density compression in front of the shock and leave a rarefaction behind.
\begin{figure}
\includegraphics[width=8.2cm]{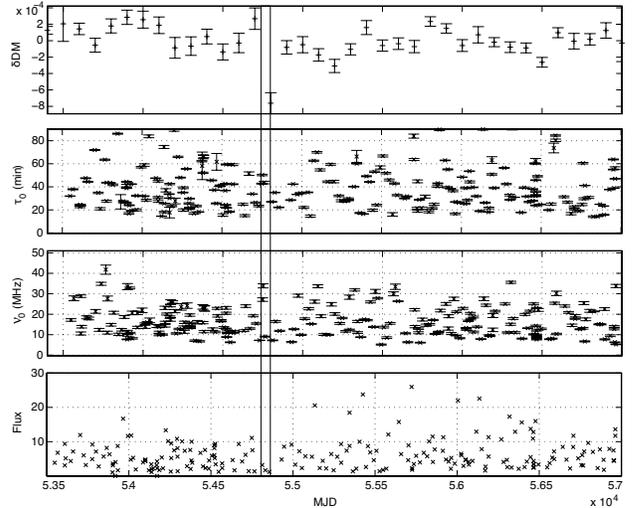}
\caption{
\label{fig2}
Observations of J1713+0747. The pinhole discussed is marked by vertical lines centered on 54800.}
\end{figure}

Nine of the 20 PPTA pulsars showed a clear linear gradient in DM(t) in the Keith et al (2013) work. However in the 3.5 yrs since the end of PPTA DR1, five of these pulsars have either lost their gradient or reversed it. An interesting example is in the observations of J1939+2134 shown in Figure 5. Here a steep negative gradient changed to an even steeper positive gradient in less than 75 days. Remarkably, the negative gradient had been observed much earlier (Ramachandran et al, 2006) and had remained sensibly constant for 20 years before the PPTA DR1.
\begin{figure}
\includegraphics[width=8.2cm]{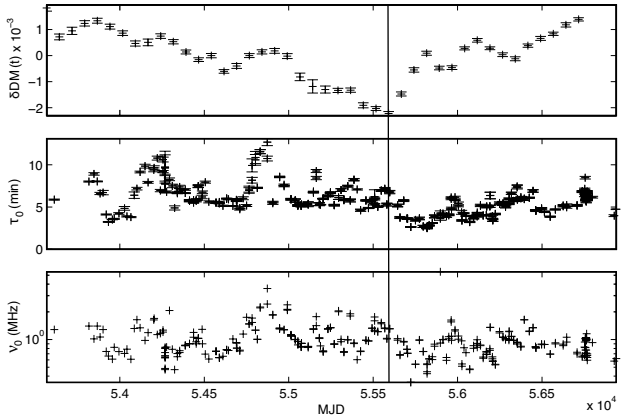}
\caption{
\label{fig2}
DM(t) observations of J1939+2134. The gradient reversal is marked by a vertical line at 55600.}
\end{figure}
Correlations between all three parameters are obvious to the eye. Some of the variation in DM(t) is caused by $V_E$ (Keith et al, 2013). As $V_P = 13.6$ km/s is unusually low it is likely that $V_{ISM}$ is also important. Although J1939+2134 has been analyzed in detail (Ramachandran et al, 2006), it would appear that further analysis, including the dynamic spectra,  as was done for the double pulsar by Rickett et al, (2014) would be useful. The correlation between $\nu_0$ and $\tau_0$ suggests that anisotropy and perhaps changes in $V_{ISM}$ should be considered. Indeed the many ``bumps'' on the DM(t) measurements for J1939+2134 may be independent clouds, each with different distance, density, turbulence, and velocity.


Of the PPTA pulsars, the only one which presently shows a very linear DM(t) gradient is J1909$-$3744. The gradient for this source could be explained by the radial velocity of the source as proposed by Cordes (2015). The radial velocity of the pulsar's white dwarf companion has been measured at -48$\pm15$ km/s (Kerkwijk, 2014). A local electron density in the vicinity of the pulsar of 5.6 cm$^{-3}$ extending over at least 100 AU would be required to explain the observed DM(t) gradient of -7.55 $\times 10^{-7} $pc cm$^{-3}$/day.

\section{Discussion}

ESEs were discovered in flux measurements of compact extra-galactic sources (AGNs) because of the strong refractive intensity variations. AGNs are too large to show diffractive intensity variations so it was unclear if the ESEs were turbulent. The implied density of the ESE is of order 10s of cm$^{-3}$, but this density is quite model dependent as it is very difficult to ``invert'' refractive fluctuations. Subsequently the phenomenon of ``parabolic arcs'' was discovered in pulsar observations. These are diffractive scintillations driven by small scale turbulence. These parabolic arc observations show many examples of discrete scattering structures of comparable size and rms density to ESEs, often many such examples in a single observation. It is tempting to identify both types of observation as ESEs, but those found in parabolic arcs are significantly more common and the observations do not provide an estimate of the mean density on larger scales.

Here we show two observations of fluctuations in DM accompanied by strong diffractive scintillations, which may provide the missing link. The DM(t) observations provide a direct measurement of the density, less model-dependent than refractive flux variations. The observations do not have sufficient signal-to-noise ratio to show parabolic arcs, but they show strong diffractive intensity variations. The observations do not show refractive intensity variations because any such variation would be obscured by the strong diffractive intensity variations. Modeling the diffractive scintillation shows that these structures, and presumably all ESEs, are very efficient at diffractive scattering because they have an ``outer scale'' of the order of AU, so they are fully turbulent on a much smaller scale than the average ISM (which has an outer scale of order pc). 

Changes in flux, group delay, or DM(t) would be observed if an ESE passes within $\theta_0$ of the line of sight to a compact radio source. However parabolic arcs from an ESE are visible when the ESE is  within $\approx 5 \theta_0$ from the pulsar (Cordes et al., 2006).  Thus a snapshot parabolic arc observation is of order 25 times more likely to find an ESE, than is a comparable observation of refraction, DM(t), or group delay.

Several of the PPTA pulsars show evidence of extended small scale variations in DM(t) persisting, in the case of J1939+2134, for 10 years or 100s of AU. It would be interesting to know more about such high density clouds. If molecular lines, or HI absorption for example, could be observed one might search for velocity structure. ESEs could be dynamic structures formed at the boundary of clouds. It would also be interesting to know if they show other molecular lines suggestive of partial ionization for example, as that would alter the nature of the turbulence. If they are related to corrugated reconnection sheets, one could search for other evidence for such reconnection sheets. In one case (Hill et al, 2005) a group of four ESEs moved together across the line of sight to a pulsar over the course of 26 days, apparently in a linear array. In another case (Brisken, et al, 2010) an image of an ESE was found to be roughly linear but well offset from the line of sight. These cases would be consistent with a sheet or a rope topology.

\acknowledgements
The Parkes radio telescope is part of the Australia Telescope, which is funded by the Commonwealth Government for operation as a National Facility managed by CSIRO.
S.O. is supported by the Alexander von Humboldt Foundation.

\end{document}